# Experimental Study of Pollutant Transfer within Dwellings

Juslin Koffi [1], Jacques Riberon [1], Ahmad Husaunndee [1], Francis Allard[2]

1. Energy Health and Environment Department
   Centre Scientifique et Technique du Bâtiment (CSTB)
   84 avenue Jean-Jaurès, BP 2, F-77447 Marne-la-Vallée Cedex 2, France.
2. LEPTAB, University of La Rochelle
   Avenue Crépeau 17042 La Rochelle Cedex, France.

*Corresponding email: juslin.koffi@cstb.fr*

## SUMMARY

A mechanical ventilation principle used in French residential buildings was tested in the experimental house of the CSTB research centre. The experiments dealt with pollutant removal efficiency of this ventilation principle, mainly with air tightness and the influence of internal doors. Tracer gas constant injection method was used to simulate the pollution source in the living room. $SF_6$ concentrations were measured in several rooms.

The results showed that the air flow routes were in agreement with the theory as long as internal doors were closed. When doors were open, the air pattern was disturbed a lot; a great quantity of the emitted pollutant was measured in the bedrooms. Besides, stack effect promotes the pollutant moving towards the bedrooms under higher indoor-outdoor air temperature differences. In addition, the results showed that if the opening of the bedroom window increases the air change rate, it does not guarantee a good indoor air quality.

## INTRODUTION

Since people spend a large amount of their time in buildings, indoor air quality is receiving increasing concern for its direct link with health, comfort, and energy consumption.

**Indoor pollutant sources**

Indoor pollutants sources can be classified into three categories: occupant sources, building materials and furnishings, and outdoor air.

People presence indoors can constitute a large range of pollutant emission depending on activities and behaviours. There are emissions from human metabolism; those are moisture, carbon dioxide, and bio-effluents. Other emissions are due to our well-being, high living standard and habits. These pollutants are humidity from cooking, body and clothes washing, and cleaning the dwelling and also combustion products like the oxides of carbon, sulphur and nitrogen.

Many of the materials used in buildings, either as structural materials or as furnishings, household chemical products, and some electric devices are sources of pollution. They release volatile organic compounds (VOC) and ozone; over two hundred VOCs have been identified in the indoor environment [1].

Indoor pollutants of interest also include airborne particles and reactive gases from traffic and industrial facilities. These may be diesel soot or the constituents of photochemical smog. The sources can also be radon which is a radioactive gas from the ground, and carbon monoxide from a garage having direct access to the dwelling.

A significant number of these pollutants, such as radon, formaldehydes, VOCs, humidity, airborne pollen and particles, can be hazardous for the occupants at high concentration level. To ensure a satisfactory indoor air quality, the concentration of indoor pollutants must be kept to safe, low risk levels. Ventilation provides one means for achieving this.

**French ventilation principle**

Three main ventilation systems can be used for this purpose:
- mechanical ventilation with local or centralised fans for exhaust, supply or balanced-ventilation with/without heat recovery;
- natural ventilation which can be performed by thermal buoyancy, or window opening;
- hybrid ventilation, a combination of both natural and mechanical ventilation.

Various strategies and principles are run to perform these systems in order to achieve good indoor air quality, energy efficiency, and comfort.

Since 1969, the French regulation on residential buildings ventilation is based on general and continuous air renewal [2,3]. The outside air enters the habitable rooms, that is the living room and bedrooms, via natural air inlets. Polluted air leaves the dwelling in service rooms which are kitchen, bathroom, shower and toilets, by mechanical exhaust vents (see figure 2). In this way, air is transferred from the higher air quality rooms to those with a lower one; the energy loss due to ventilation is thus reduced because heated air coming from the habitable rooms is used for ventilating the service rooms. An other advantage of such a regulation is that, whatever the weather may be, the flow of ventilation remains constant and known in each service room [3]. This ventilation principle is so-called here CVC principle as "*Central Ventilation by air Circulation through the dwelling*" or central exhaust mechanical ventilation. It is allowed in all the residences, single and multi-family houses, by the French ventilation regulation [2,5,6].

Internal air movements are governed by two main factors [7]:
- natural driven forces: wind, stack effects due to air density difference between rooms;
- and mechanical forces driven by operating fans.

In some cases, at high concentration difference of humidity or gaseous pollutants, the flow can be also governed by the concentration gradient. In mechanical systems, they result from both natural and mechanical effects. However, the latter should be prevalent on stack effect and infiltrations in order to impose internal flows direction, and then achieve mechanical ventilation purposes [7].

The objective of the current study is to test out, by experimental means, the CVC ventilation principle in order to check which of the above parameters are influential on the internal air flows, so on pollutant transport, and how they can effectively affect its operation. The lifted question is therefore to known whether a pollutant coming from the living room can reach a bedroom on the first floor according to a number of conditions.

## METHODS

### The experimental house

Experiments on CVC mechanical ventilation principle had been carried out in the CSTB experimental house MARIA (Mechanized house for Advanced Research on Indoor Air) [9], for various conditions with regard to windows and internal doors. MARIA is a three-level house is located at the CSTB research centre in Marne-la-Vallée, near Paris. It has four bedrooms, a shower, and a bathroom on the first floor. The living, the kitchen, and the toilets are situated on the ground floor. There is a garage on the underground floor.

Purpose-built orifices are located in the facades in order to simulate different levels of envelope air leakage in the house. Experimental evaluation performed gave a residual air leakage value of 1.2 m$^3$/h/m$^2$, and when 30 apertures were open, air leakage was 2.2 m$^3$/h/m$^2$. Further information about these results is available in [10].

### Experiments

Tracer gas constant injection method was used to simulate a pollutant emission in the living room. Sulphur hexafluoride, $SF_6$, was so continuously released during 3 hours at 1.5 ml/s flow rate to represent occupant presence or activity. Concentrations were continuously measured at six different points within the dwelling (figure 2): measurement sensors were placed in the living, the kitchen, the bathroom, the shower, the hall, and in bedroom 3. The tracer gas was considered absent outside. Pollutant concentration in the bedroom aimed to show if internal flows followed the theory of the CVC principle. After the injection, decay method was used to evaluate air change rate.

These measurements were done with a six-point sampler for pumping air from measurement points, and a multi-gas monitor that calculated pollutant concentrations. All devices were placed at the underground floor, and the analysed air from the monitor was rejected outside and far from the building to avoid additional source.

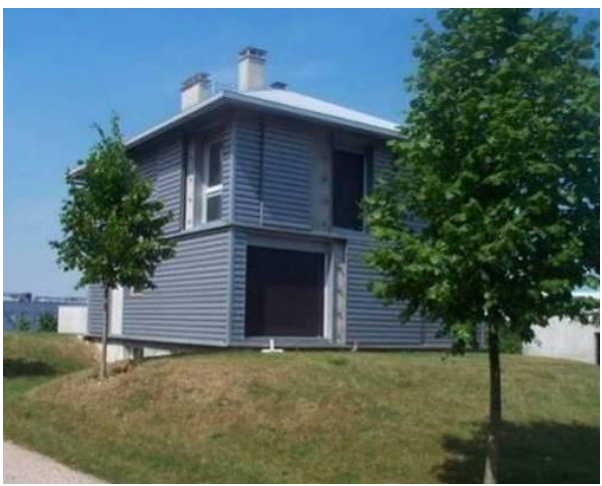 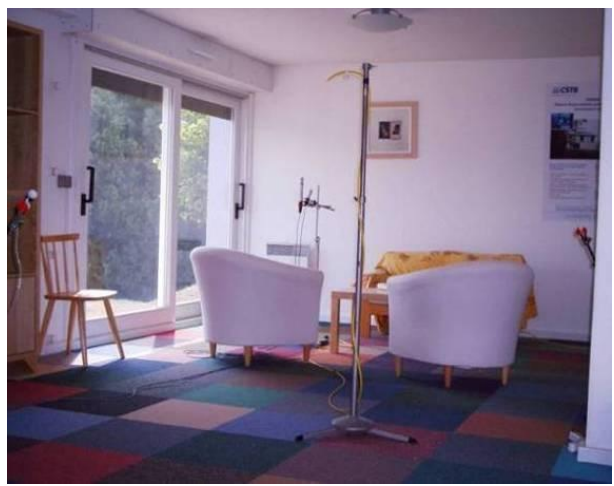

a) b)

Figure 1: The CSTB experimental house MARIA.
a) Outside view, b) Pollutant injection and measurement in the living room.

**Cases studied**

During the experiments, all external doors, internal doors of service rooms and windows were kept closed. We then studied many situations with regard to internal doors of living and bedrooms, air leakage openings, and windows:

- Case 1: internal doors and air leakage orifices were closed;
- Case 2: internal doors were closed and 30 orifices were open;
- Case 3: internal doors were open and orifices were closed;
- Case 4: internal doors and 30 orifices were open;
- Case 5: internal doors and 30 orifices were open, and exceptionally the window of bedroom 3 was open too.

To run the CVC principle in MARIA, the house was equipped with mechanical exhaust ventilation and natural air inlets in the habitable rooms. The configuration was as follows (fig. 2):

- continuous extraction in the kitchen at about 60 m$^3$/h flowrate;
- continuous extraction in both shower and bathroom at a mean of 33 m$^3$/h flowrate per room;
- mechanical extraction was stopped in the toilets (in order to make some equilibrate flows between the two levels of the building).

In each case, outdoor and rooms temperatures, and wind speed were measured by the mean of the CSTB meteorological station. Numbers of tests have been carried out in order to check out the repetitiveness of the results. The following part deals with the obtained results and their analysis.

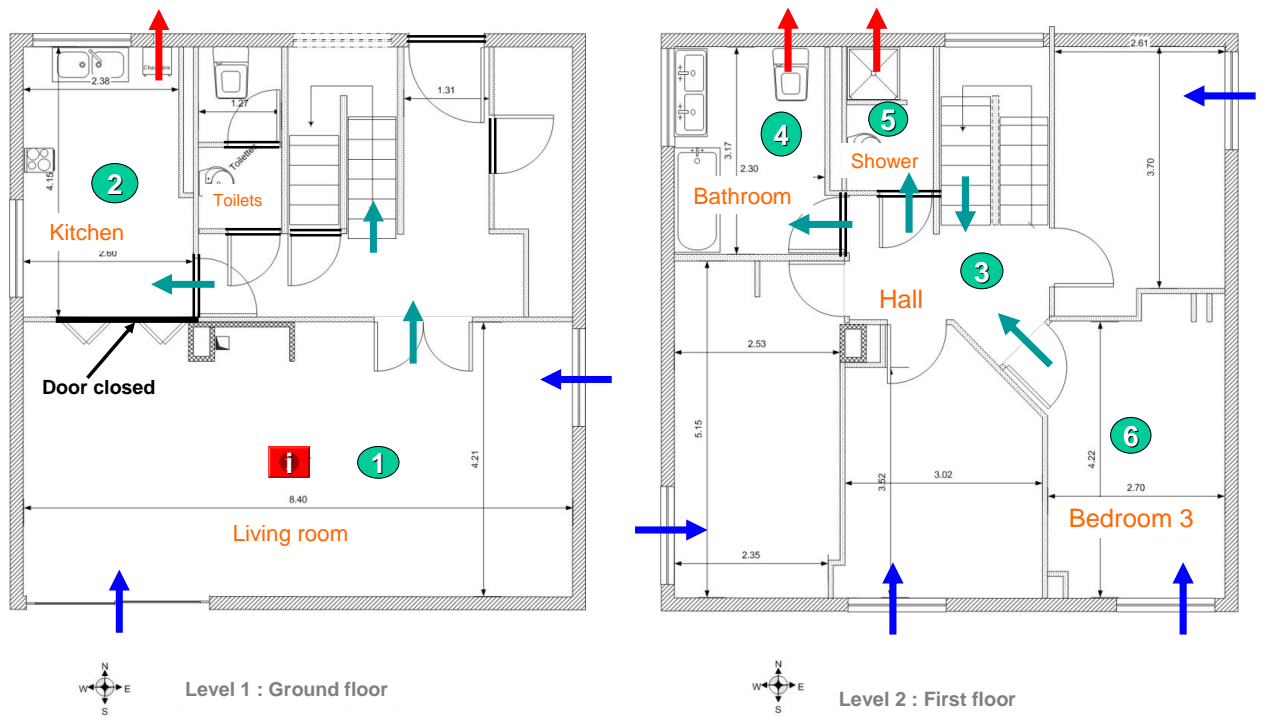

Figure 2: Pollutant emission and the measurement points within the experimental house.

# RESULTS

Case 1: The results of experiment *case 1*, where all doors and apertures are kept closed, are illustrated on figure 3a) where indoor $SF_6$ concentrations are shown according to time. As it can be seen on the picture, concentration level in the living is very different from those of the other rooms. One can note some fluctuations on the curve representing pollutant concentration level in the living room; in the other rooms, the curves evolve in a progressive way. Following these observations, an averaged value of the release zone concentration was considered for the analyses, and the maximum level for the other rooms.

In this way, about 8 ppm of $SF_6$ was measured in bedroom 3 and 20 ppm in the other rooms of the first floor, while the averaged concentration was round about 82 ppm in the living during the last hour of emission period, and roughly 15 ppm in the kitchen. In this case, air flows seemed seems to be well driven by mechanical ventilation forces as soon as in two other tests bedroom 3 concentrations dropped to 1 ppm while values grew or remained almost the same in the rest of the building (figure 3b)).

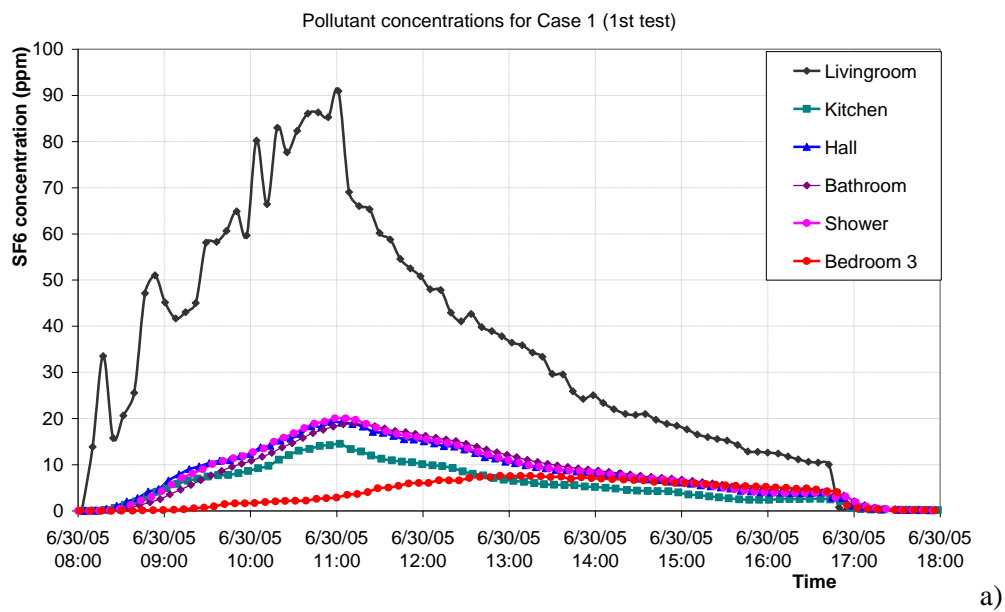

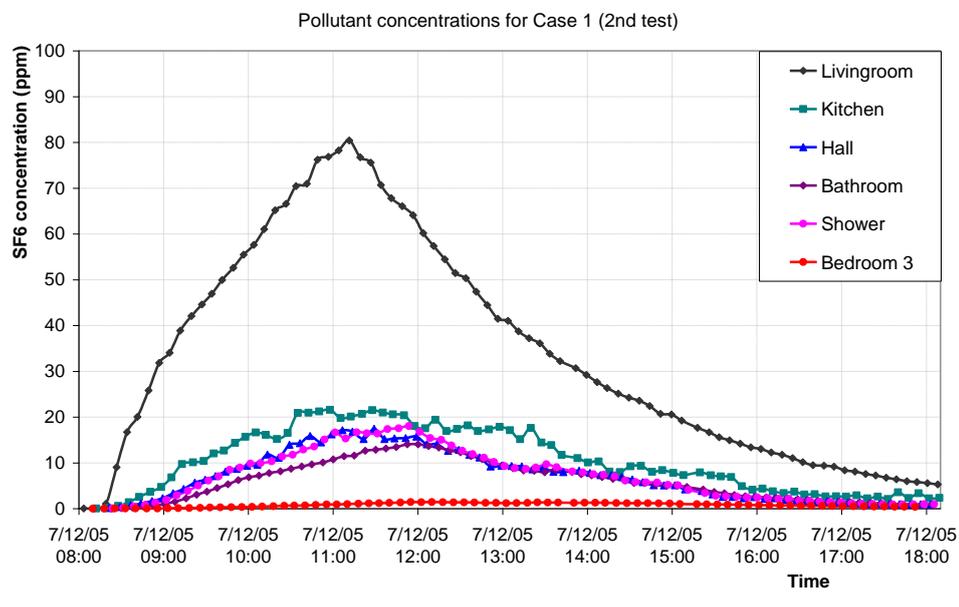

Figure 3: Pollutant concentrations within MARIA for experiment case 1. a) Test 1, b) Test 2.

Case 2: When the orifices were open, the concentration in the living room decreased to about 64 ppm. But, 7 ppm was measured in bedroom 3, 14 ppm in the kitchen, and up to 20 ppm in the service rooms at the first floor. Like in case 1, two complementary tests gave lower than 1 ppm of $SF_6$ concentration in bedroom 3. The effects of air leakage openings was visible on living room concentration but not in the bedroom.

Case 3 and Case 4: The opening of internal doors in *case 3* provoked a huge growth of concentration in rooms at the first floor (figure 4). In bedroom 3, as well as in the bathroom and the shower, it was measured 25 ppm, more than three times the former bedroom levels. The same range of concentrations was found when the air leakage orifices were open in addition to the doors, that is between 21 and 24 ppm (see figure 5). In these two cases, pollutant concentrations in the living room were 53 then 42 ppm; respectively 15 and 11 ppm were found in the kitchen.

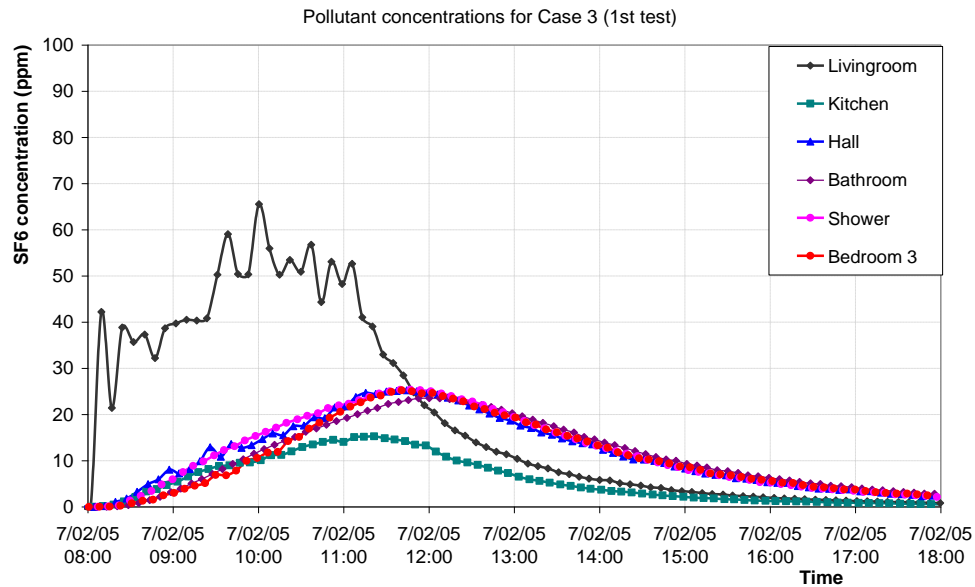

Figure 4: Pollutant concentrations within MARIA for experiment *case 3*.

Case 5: In the final case, when internal doors of main rooms, air leakage openings and window in bedroom 3 were open, pollutant concentration within the dwelling was greatly diluted (figure 6). Apart from the source room where it was measured 41 ppm and the hall with 16 ppm, $SF_6$ levels were lower than 10 ppm in the whole housing.

In short, by looking through all results, four categories of rooms can be considered according to measured $SF_6$ concentrations:
- the living room: averaged concentrations were the highest and ranged from 38 to 82 ppm;
- the kitchen: concentration ranged from 6 to 15 ppm;
- the hall and the service rooms at first floor where concentrations varied from 6 to 26 ppm. Concentrations in bathroom and shower were still equal as mechanical exhaust rate was the same;
- the bedroom: concentrations ranged from 0 to 25 ppm.

Table 1 shows the results of all tests. In this table, concentrations in the bedroom are compared to those in the release room by mean of the ratio R. This ratio is used to assess the effectiveness CVC ventilation principle and is simply defined as:

$$R = \frac{bedroom\ concentration}{living\text{-}room\ concentration}. \quad (1)$$

The lower *R* is, the better the ventilation principle works.

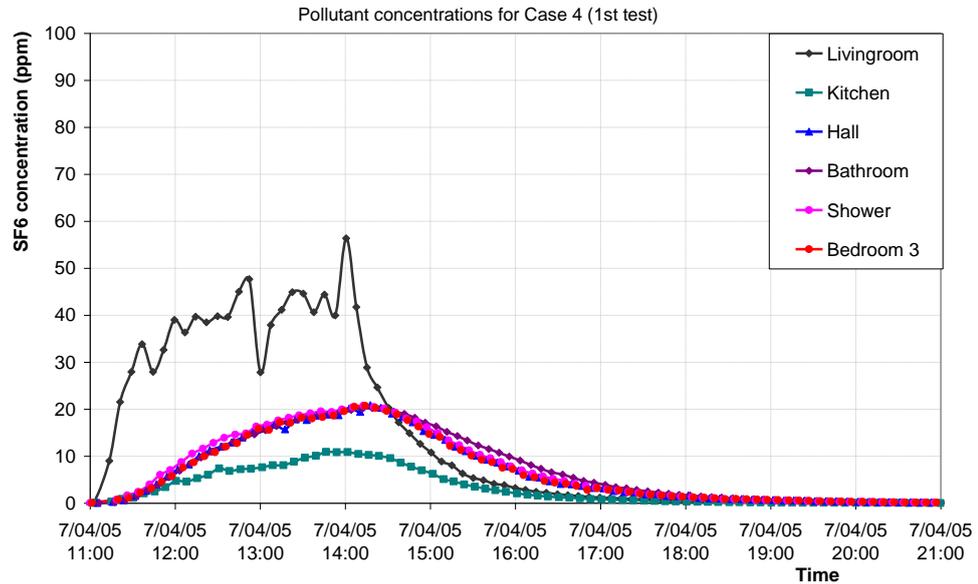

Figure 5: Pollutant concentrations within MARIA for experiment *case 4*.

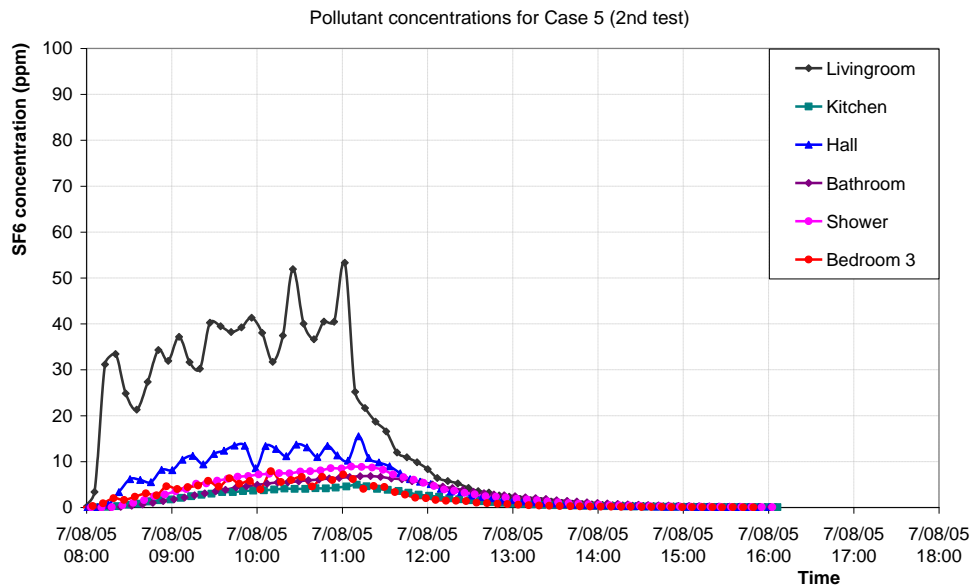

Figure 6: Pollutant concentrations within MARIA for experiment *case 5*.

In *cases 1* and *2,* the value of R did not exceed 0.11, that is bedroom concentration was under 11% of living room concentration. The CVC principle can be considered well as operating when doors are closed. R is about 0.50 for open door configurations, where a great quantity of pollutant where measured in the experimented bedroom. The depression due to mechanical ventilation becomes insufficient to totally govern the flows between rooms.

When analysing table 1, inside-outside temperature difference ($\Delta T_{IO}$) effect on pollutant concentration in bedroom 3 is clearly notable. In *case 1*, when $\Delta T_{IO}$ is 9°C, concentration is 8 ppm; for $\Delta T_{IO}$ below 3°C, concentration is only 1 ppm maximum. A similar observation can be made in *case 2*. It therefore seems that stack effect also promotes the pollutant moving towards the bedrooms, thus appearing that an important parameter of internal air flows.

Table 1: Pollutant concentrations and the ratio R, weather conditions and air renewal rate.

|  | | Living room (ppm) | Bedroom (ppm) | R | $T_{Bed\ 3}$ (°C) | $T_{ext}$ (°C) | Wind (m/s) | N (vol/h) |
|---|---|---|---|---|---|---|---|---|
| Case 1 | Test 1 | 82 | 8 | 0,10 | 29 | 20 | 2,4 | 0,35 |
|  | Test 2 | 72 | 1 | 0,01 | 24 | 24 | 4 | 0,40 |
|  | Test 3 | 64 | 0 | 0 | 26 | 24 | 2,1 | 0,47 |
| Case 2 | Test 1 | 64 | 7 | 0,11 | 27 | 19 | 2,4 | 0,52 |
|  | Test 2 | 62 | 1 | 0,02 | 22 | 25 | 3,6 | 0,59 |
|  | Test 3 | 55 | 2 | 0,04 | 26 | 25 | 1,7 | 0,43 |
| Case 3 | Test 1 | 53 | 25 | 0,47 | 26 | 24 | 2,1 | 0,57 |
|  | Test 2 | 55 | 22 | 0,40 | 25 | 25 | 3,2 | 0,43 |
| Case 4 | Test 1 | 42 | 21 | 0,50 | 26 | 19 | 3,4 | 1,08 |
|  | Test 2 | 40 | 22 | 0,54 | 25 | 19 | 2,6 | 0,92 |
| Case 5 | Test 1 | 38 | 6 | 0,16 | 20 | 16 | 3,2 | 1,36 |
|  | Test 2 | 41 | 8 | 0,19 | 20 | 16 | 2,8 | 1,32 |

Infiltration due to orifices opening and wind diluted pollutant concentration and increased air renewal rate (N) in the living room. Factor N, evaluated by the decay method, averagely ranged from 0.40 volume per hour in *case 1* to 0.50 vol/h in *case 2*. When doors were open in addition to holes, air change was twice that value. The opening of air leakage orifices resulted in a slight decrease of concentration in the living room. Ratio R was not affected when passing from *case1* to *case 2*; moreover, if one considers that R values higher than 0.10 in these cases were only due to stack effect, infiltration influence on R declines under 4%. But, one can underline a slight raise of R from *case 3* (0.40-0.47) to *case 4* (0.50-0.54). Then, pollutant concentration in bedroom 3 seemed to be sensitive to orifices opening, thus to infiltration, only when doors were open.

Window airing increased air change rate (1.32-1.36 vol/h) and diluted concentration in the whole house. Its effects were prevailing to mechanical ventilation. The values of R (0.16-0.19) and concentrations showed that pollutant removal in this way was not too efficient.

## DISCUSSION

The tests performed in the experimental house MARIA upon French ventilation principle gave the following results. The air flow routes were in agreement with the CVC principle as long as the internal doors were closed. The opening of these doors disturbed a lot the air pattern since great quantity of the emitted pollutant was measured in the experimented bedroom. Beside that, stack effect promoted the pollutant moving towards the bedrooms for higher indoor-outdoor air temperature differences. In addition, the results also showed that if the opening of the bedroom window increased the air change rate, it did not guarantee a good indoor air quality; and infiltration due to the purpose-built orifices mostly affected concentration in the emission room.

Nevertheless, it seems important to recall some of the simplifications made during the experiments: direct flow between living and kitchen was not permitted, the door was made air tight, and extraction in toilets was stooped. Furthermore the openings were equally distributed between the levels, and as well as possible between rooms, in order to represent a globally leaky building envelope. Other configurations such as a more localised distribution of orifices can lead to different results. Concerning the method itself, as concentrations never reached steady state, it would be interesting to see the influence of such a situation.

Despite all, one must note that the CVC principle works well. However, it remains sensitive to numbers of parameters, mainly the opening of the doors which is related to occupants' behaviour. It then seems important to pay more attention to these disruptive factors in order to maintain the well operating conditions. Otherwise, it should be interesting to think about other ventilation strategies able to achieve the purposes of good indoor air quality, energy efficiency, and comfort.

## ACKNOWLEDGMENT

The authors wish to thank Professor Nachida Bourabaa from ENSIAME Engineering School, University of Valenciennes (France) for her assistance.

## REFERENCES

[1] S. K. Brown, M. R. Sim, M. J. Abramson, and C. N. Gary, *Concentration of volatile organic compounds indoor air-A review*. Indoor Air, 1994, 123-134.

[2] Journal Officiel, *Arrêté du 24 mars 1982 (urbanisme et logement, énergie, santé) modifié par arrêté du 28 octobre 1983, Dispositions relatives à l'aération des logements.* Journal Officiel de La République Française, 27 mars 1982 et 15 novembre 1983.

[3] J. Ribéron, J. G. Villenave, J.-R. Millet, *Evaluation of mechanical domestic ventilation systems: French approach.* International conference Indoor Air'92, Helsinki $4^{th}$-$8^{th}$ jusly 1993.

[4] J. G. Villenave, J. Ribéron, J.-R. Millet, *French ventilation in dwellings.* Air Infiltration Review, vol. 14 n°4, 1993.

[5] B. Brogat, J. Fontan, P. Lanchon, J-R. Millet, C. Skoda-Schmoll, J-G. Villenave, *Ventilation dans les bâtiments collectifs d'habitation existants.* Cahiers du CSTB, livraison 412, septembre 2000.

[6] J. Ribéron, *Les différents systèmes de ventilation et l'influence de la perméabilité du bâtiment.* Formation Professionnelle Continue 22-23-24 mai 1996 - Ventilation de bâtiments, Document de référence - Centre Scientifique et Technique du Bâtiment, 1996.

[7] D. Etheridge, M. Sandberg, *Building Ventilation: Theory and Measurement*. British Gas plc, Gas Research Centre, Loughborough, UK. Royal Institute of Technology, Byggd, Gävle, Sweden. John Wiley & Sons, 1996.

[8] M. A. Lamrani, *Transferts thermiques et éérauliques à l'intérieur des bâtiments.* Thèse de doctorat soutenue le 27 mars 1987, Université de Nice – CSTB Sophia-Antipolis, 165p.

[9] J. Ribéron, P. O'Kelly, *MARIA: an experimental tool at the service of indoor air quality in housing sector.* Indoor Air'2002, International Conference Proceedings, Monterey (Canada). June 30–July 5, 2002.

[10] J. Koffi, *Etude expérimentale du transport de polluant entre pièces d'un logement.* Rapport de stage de fin de cycle Ingénieur et de Master recherche, document CSTB n° DDD-DE-VAI 05-061R. 2005.